%% file: main.tex
\documentclass[acmsmall,screen]{acmart}

\usepackage{amsmath}
\usepackage{dsfont}
\usepackage{algorithmic}

\usepackage{xspace}
\usepackage{multirow}
\usepackage{listings}
\usepackage{booktabs}
\usepackage{tabularx}
\usepackage{threeparttable}
\usepackage{array}
\usepackage{enumitem}

\input{sections/0x99command}

\makeatletter
\def\@authorsaddresses{}
\makeatother

\begin{document}

\title{LLM-based Code Auditing by Inferring and Evaluating Developer Assumptions}

\author{Junming Liu}
\email{junming.liu@mpi-sp.org}
\affiliation{%
  \institution{MPI-SP}
  \country{Germany}
}

\author{Marcel Böhme}
\email{marcel.boehme@acm.org}
\affiliation{%
  \institution{CISPA}
  \country{Germany}
}

\input{sections/0x00Abstract}

\maketitle

\input{sections/0x10Introduction}

\input{sections/0x20Approach}
\input{sections/0x30Metric}
\input{sections/0x40Evaluation}
\input{sections/0x60RelatedWork}
\input{sections/0x50Discussion}
\input{sections/0x70Conclusion}


\newpage
\bibliographystyle{ACM-Reference-Format}
\bibliography{Ref}

\end{document}

%% file: sections/0x99command.tex
\newcommand{\beliefanalyzer}{\textsc{Creed}\xspace}

\newcommand{\primevulexp}{PrimeVul-CoT\textsubscript{exp}\xspace}

\newcolumntype{C}{>{\centering\arraybackslash}X}
\newcolumntype{L}{>{\raggedright\arraybackslash}X}
\newcolumntype{M}[1]{>{\raggedright\arraybackslash}m{#1}}
\newcolumntype{P}[1]{>{\raggedright\arraybackslash}p{#1}}

\newif\ifcommentcond
\commentcondtrue 
\newcounter{ljm}

\newcounter{marcel}

\newcounter{todo}

\definecolor{codebg}{RGB}{248,249,251}
\definecolor{codeframe}{RGB}{215,220,228}
\definecolor{kwcolor}{RGB}{215,58,73}
\definecolor{strcolor}{RGB}{3,136,3}
\definecolor{commentcolor}{RGB}{106,115,125}
\definecolor{numcolor}{RGB}{90,100,115}
\definecolor{funccolor}{RGB}{111,66,193}

\definecolor{promptbg}{RGB}{248,248,248}
\definecolor{promptframe}{RGB}{220,220,220}

\lstdefinelanguage{TypeScript}{
  keywords={
    export,async,function,const,let,var,await,return,if,else,try,catch,
    true,false,boolean,string,Promise
  },
  keywordstyle=\color{kwcolor}\bfseries,
  sensitive=true,
  comment=[l]{//},
  morecomment=[s]{/*}{*/},
  string=[b]",
  morestring=[b]',
}

\usepackage{lstlinebgrd}
\definecolor{hlline}{RGB}{255,255,200}

\lstdefinestyle{base}{
  language=TypeScript,
  backgroundcolor=\color{codebg},
  basicstyle=\ttfamily\footnotesize,
  keywordstyle=\color{black},
  stringstyle=\color{black},
  commentstyle=\color{black},
  numberstyle=\scriptsize\color{numcolor},
  numbers=left,
  firstnumber=1,
  stepnumber=1,
  numbersep=8pt,
  numberblanklines=true,
  frame=single,
  framerule=0.4pt,
  rulecolor=\color{codeframe},
  framesep=6pt,
  xleftmargin=18pt,
  xrightmargin=6pt,
  breaklines=true,
  breakatwhitespace=false,
  postbreak=\mbox{\textcolor{commentcolor}{$\hookrightarrow$}\space},
  showstringspaces=false,
  keepspaces=true,
  columns=fullflexible,
  tabsize=2,
  captionpos=b,
  aboveskip=0.8em,
  belowskip=0.8em
}

\definecolor{promptbg}{RGB}{249,249,250}
\definecolor{promptframe}{RGB}{210,212,218}
\definecolor{promptkey}{RGB}{40,80,143}

\lstdefinestyle{prompt}{
    backgroundcolor=\color{promptbg},
    basicstyle=\small\ttfamily,
    frame=single,
    framerule=0.5pt,
    rulecolor=\color{promptframe},
    framesep=8pt,
    breaklines=true,
    breakindent=0pt,
    columns=fullflexible,
    keepspaces=true,
    showstringspaces=false,
    aboveskip=10pt,
    belowskip=6pt,
    xleftmargin=6pt,
    xrightmargin=6pt,
    captionpos=b,
    abovecaptionskip=8pt,
}

\usepackage{tcolorbox}
\newcommand{\result}[1]{%
\begin{tcolorbox}[colframe=black,boxrule=0.5pt,arc=4pt,
      left=6pt,right=6pt,top=6pt,bottom=6pt,boxsep=0pt,width=\columnwidth]%
      {\emph{#1}}
\end{tcolorbox}%
}

%% file: sections/0x00Abstract.tex
\begin{abstract}
When we see a null pointer check, we know that the developer believes that the pointer may \emph{not} be initialized. If we see the same pointer later used without a null pointer check, we may consider this a violation of the developer's belief and ultimately a bug. In this paper, we develop such \emph{a belief-guided program analysis for vulnerability detection} (VD). 
For example, when an authentication function returns the same error for both, a wrong user name and a wrong password, a human auditor may infer that the developer believes that both cases are indistinguishable, mitigating a user enumeration attack. This belief is reasonable, as it follows standard security practice. Yet, a function call inducing a difference in response time for both cases, contradicts this belief. This contradiction uncovers a timing side-channel that syntactic analysis tools would miss.
Existing static analysis tools struggle with this type of flaws because they require them to be expressible by syntactic patterns.
Our belief-guided approach allows us to detect such domain-specific, high-level semantic security flaws that have so far remained elusive.

Given a function, our approach, \beliefanalyzer,~infers the developer's security-related beliefs, evaluates how reasonable each one is, and derives the conditions under which it is violated. In contrast to existing ML-based VD techniques,
\beliefanalyzer \emph{explains} the root cause, the symptoms, and the conditions under which each issue is exploitable.
To overcome the context-dependency problem in the \emph{evaluation}, we introduce two measures of the performance of explanation-based VD: \textit{recall} quantifies how effectively an approach correctly explains a known vulnerability, and \textit{overhead} quantifies the effort of inspecting its reports.
On a recent real-world vulnerability benchmark, \beliefanalyzer~detects both semantic and syntactic vulnerabilities more effectively than the existing explanation-based VD tools, achieving the best trade-off between effectiveness and effort.
\end{abstract}

\keywords{
Vulnerability detection, static analysis, large language models
}

%% file: sections/0x10Introduction.tex
\section{Introduction}
Many types of security flaws require domain-specific, high-level semantic code understanding. For instance, in our random sample of 50 recent vulnerabilities collected by CVEfixes~\cite{dataset:cvefixes}, developers forgot to consider a timing side-channel when mitigating user enumeration (CVE-2026-25222), forgot to check object-level authorization before canceling a payment (CVE-2026-2010), 
forgot to limit a path name only to the allowed directory (CVE-2026-28427), or forgot to verify a cryptographic signature (CVE-2026-22866). 
How can we discover these automatically and at scale?

To detect such semantic security flaws manually, code auditors follow a process of identifying developer beliefs and locations where they are contradicted. Cognitive studies suggest~\cite{humanauditor:cognitivecode, humanauditor:hypothesis, humanauditor:codereview}: Developers carry certain \emph{beliefs} (i.e., implicit assumptions) that can be identified in the source code. Auditors iteratively infer the developer's beliefs about the program's behavior, form expectations based on their domain knowledge, and identify bugs where the implementation deviates from these expectations.

Existing attempts to automate this \emph{belief-guided analysis} are based on traditional static analysis \cite{static:deviant,belief:specification,belief:prminer}.
Beliefs and their violations are expressed and checked as general patterns on static code representations. For instance, a null-pointer check guarding a dereference implies the \emph{belief} that the pointer may be null. The existence of a path along which that pointer is (not reassigned but) dereferenced \emph{without} a null-pointer check is \emph{inconsistent} with this belief. An analysis to detect such paths might use the program dependence graph (PDG).

However, identifying general beliefs a priori and encoding them as patterns within a syntactic representation of the source code (e.g., PDG) fundamentally limits the types of bugs that can be found. Most of the recent security flaws, we have encountered, violate \emph{domain-specific, high-level beliefs} that cannot be expressed as such patterns. For instance, Listing~\ref{lst:func-signin} shows the function patched to fix CVE-2026-25222. How can we expect a traditional static analysis to work for the \texttt{TypeScript} language or to figure out, this function handles user authentication? How can we expect that it identifies "invcreds" as an error code or that the developer believes that returning the same error code after entering the wrong user or password would not leak that the user existed (to mitigate a user enumeration attack)? These are broad semantic facts specific to the authentication domain that can hardly be derived, e.g., from the (language-specific) PDG.

\input{tables/FuncSignin.tex}

In this paper, we propose a \emph{belief-guided, language-agnostic, semantic static analysis for vulnerability detection} that implements a general belief-guided approach using large language models (LLMs).
Our approach, \beliefanalyzer, identifies what the given function tries to achieve (action), which assumptions the developer made when implementing this action (belief), whether these assumptions are reasonable (sanity), and whether there are conditions under which they are violated (violation). The bugs thus detected can be as low-level and general as missing null pointer checks and as high-level and specific as unintended timing side channels.

In Listing~\ref{lst:func-signin}, the function returns the same error code "invcreds" if the user name or if the password is incorrect (action). The corresponding belief may be that a generic error response makes the two failure modes indistinguishable to attackers (belief). Doing so is standard security practice and prevents user enumeration (sanity). If the response time differs significantly between both failure modes, e.g., due to the additional function call, an attacker could still enumerate valid users (violation). This is CVE-2026-25222.

How do we evaluate whether our belief-guided approach is effective? Unlike many existing vulnerability detection techniques, our approach does not classify a function as vulnerable or not, but produces an \emph{explanation} of why and under which conditions a function may cause a vulnerability. In fact, casting machine learning for vulnerability detection as a function-level binary classification problem is invalid~\cite{metric:niklasbench} because in most cases whether or not a function causes a vulnerability depends on the context in which this function is called. This context-dependency problem is alleviated in our problem statement because i) we offer an explanation that a human auditor can check, and ii) the explanation is context-conditional.

We suggest evaluating the \emph{recall} of our approach as the proportion of vulnerabilities where our approach produces an \emph{explanation that is consistent with the available evidence} (i.e., CVE description, patch commit message, and patch diff code).
We cannot compute the \emph{precision} of our approach automatically since many of \beliefanalyzer's reports, beyond the one we expect to find, seem reasonable to us. Nevertheless, to get a handle on precision, we manually analyze the first report of each function.
Moreover, we evaluate the \emph{overhead} of our approach as the number of reports per expected vulnerability.

We evaluate \beliefanalyzer~on 50 recent CVEs spanning 10 programming languages, 40 of which are semantic flaws. \beliefanalyzer~achieves the highest recall (0.78), correctly identifying and explaining the root cause of 39 of 50 vulnerabilities. Compared with the strongest baseline, it improves recall by 10 percentage points while lowering overhead from 11.3 to 7.9, about 30\%. Even under a tight inspection budget, it surpasses the recall any baseline reaches with unlimited effort. It leads on both memory-related and semantic vulnerabilities, including 15 of 21 cases that cannot be expressed as general static-analysis patterns.

\vspace{0.1cm}
\noindent
In summary, we make the following \emph{contributions}.\vspace{-0.1cm}
\begin{itemize}[leftmargin=0.4cm]
    \item We propose \beliefanalyzer, a belief-guided, language-agnostic, semantic static analysis for vulnerability detection.
    \item We evaluate explanation-based detection with \textit{recall} and \textit{overhead}, addressing the context-dependency problem in function-level vulnerability detection.
    \item We demonstrate the effectiveness of \beliefanalyzer~on both semantic and memory-related vulnerabilities.
\end{itemize}

%% file: tables/FuncSignin.tex
\begin{lstlisting}[
  style=base,
  escapeinside={(*@}{@*)},
  linebackgroundcolor={%
  \ifnum\value{lstnumber}=5\color{hlline}%
  \else\ifnum\value{lstnumber}=11\color{hlline}%
  \else\color{codebg}%
  \fi\fi},
  caption={User enumeration vulnerability in PolarLearn (CVE-2026-25222)},
  label={lst:func-signin}
]
export async function signInCredentials(
    email: string, password: string): ... {
  ...
  const user = await prisma.user.findUnique({ where: { email },});
  if (!user) return "invcreds";
  ...
  (*@\textcolor{red}{const hashedPassword = await hashPassword(password, user.salt);}@*)
  if (user.password === hashedPassword) {
    await createSession(user.id);
    return true;
  } else return ("invcreds");
  ...
}
\end{lstlisting}

%% file: sections/0x20Approach.tex
\section{Belief-Guided Static Analysis for Vulnerability Detection via  Language Models}

\begin{figure}[tbh]
    \centering
    \includegraphics[width=.7\linewidth]{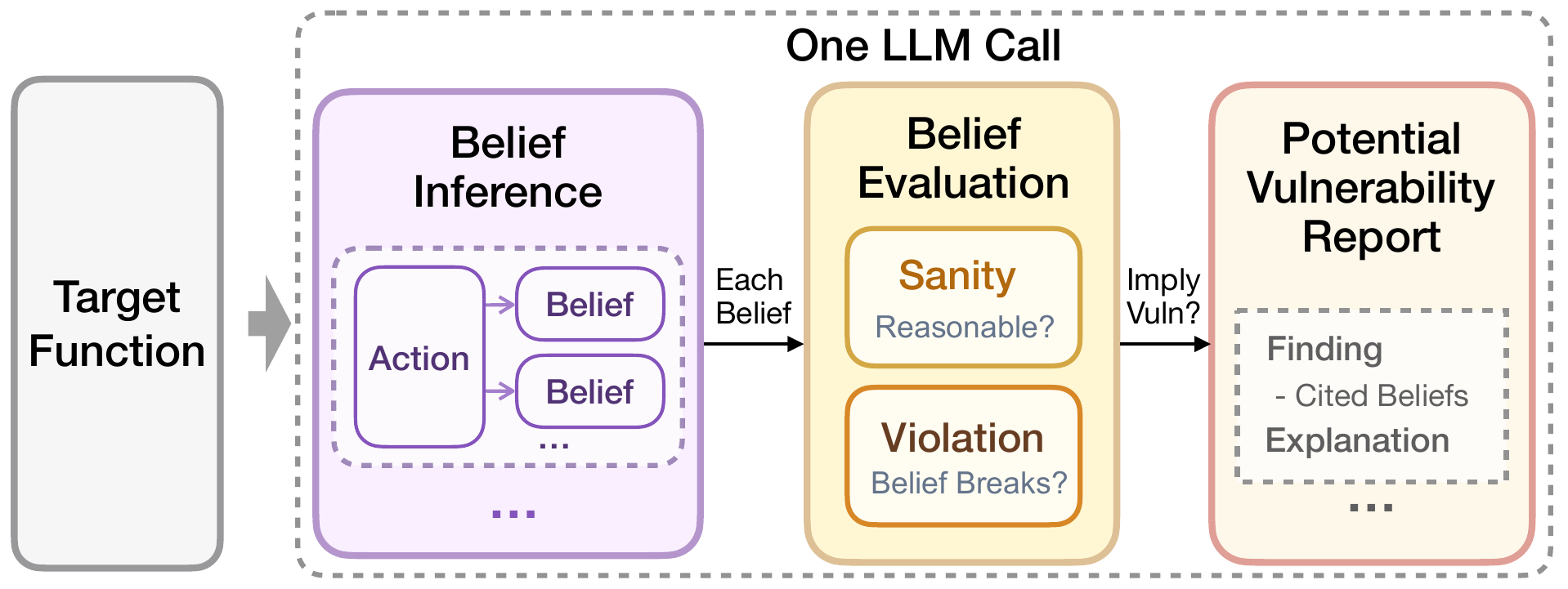}
    \caption{Overview of \beliefanalyzer.}
    \label{fig:overview}
\end{figure}

\subsection{Overview}

Figure~\ref{fig:overview} shows the overall workflow of \beliefanalyzer.
Given a function, \beliefanalyzer~makes explicit the developer's implicit beliefs about the action implemented in the function, evaluates them, and produces a \textit{vulnerability report}.
Specifically, \beliefanalyzer~proceeds in three stages.
During \textit{belief inference} (Section~\ref{sec:inference}), it identifies the actions the code performs and infers the beliefs the developer holds when performing them.
During \textit{belief evaluation} (Section~\ref{sec:evaluation}), it evaluates each belief through two complementary checks: a sanity check that determines whether the belief is reasonable (i.e., consistent with established security standards), and a violation check that derives the conditions under which the belief may no longer hold.
During \textit{vulnerability reporting} (Section~\ref{sec:identification}), \beliefanalyzer produces a report for each finding that explains the nature of the potential vulnerability, the root cause and the context under which it can be exploited.
\beliefanalyzer~realizes this workflow with a single prompt (Fig.~\ref{lst:prompt}) that guides an LLM through the three stages and returns a structured result.

\input{tables/Prompt}

\input{tables/FuncCancel.tex}

\textbf{Examples.}
Table~\ref{tab:belief-example} walks through \beliefanalyzer~on two functions, which together illustrate the two ways a belief can uncover a potential vulnerability.
Listing~\ref{lst:func-signin} shows our first example from the introduction, a TypeScript function called \texttt{signInCredentials} fixed to patch CVE-2026-25222.
Listing~\ref{lst:func-cancel} shows our second example, a Java function called \texttt{cancel} fixed to patch CVE-2026-2010.

For the \texttt{signInCredentials} function (Listing~\ref{lst:func-signin}; Table~\ref{tab:belief-example}.left), \beliefanalyzer identifies the \emph{action}, which the function implements, as authentication, which returns the same generic error for both non-existent users and wrong passwords. 
It then infers the developer's \emph{belief} that the generic error makes the two cases indistinguishable which it evaluates as \emph{reasonable}. However, when analyzing whether there are conditions under which this \emph{belief is violated}, \beliefanalyzer finds that the additional call to \texttt{hashPassword} might induce a difference in response times causing a timing side-channel and thus facilitating the user enumeration vulnerability that the developer meant to prevent.

\input{tables/BeliefExample.tex}

For the \texttt{cancel} function (Listing~\ref{lst:func-cancel}, Table~\ref{tab:belief-example}.right), \beliefanalyzer identifies the \emph{action}, which the function implements, as canceling a pending payment, which checks only the \texttt{siteId} and the payment status before canceling a payment. It then infers the developer's \emph{belief} that a matching \texttt{siteId} is sufficient to authorize the cancellation of the payment. However, \beliefanalyzer finds this \emph{unreasonable} on its own, since belonging to the same site does not imply the right to cancel that particular payment.

The two cases show that \beliefanalyzer~reports potential vulnerabilities both from a belief that is unreasonable and from a reasonable belief whose violation condition can be triggered. Both functions are written in different languages.

\subsection{Belief Inference}
\label{sec:inference}

A \textit{belief} is an assumption the developer holds when implementing the code.
Beliefs cannot be observed directly. Instead, they are inferred from the observable behaviors the code performs, which we refer to as \textit{actions}.

\textbf{Action Identification.}
A function can be understood as a single action in itself (like the \texttt{signInCredential} function implementing authentication). Yet, a function also comprises multiple finer-grained actions that together achieve its overall functionality (like the various checks during authentication). 
Fixing a specific granularity for identifying actions is therefore insufficient: low-level actions and the function's overall purpose expose different classes of potential vulnerabilities.
\beliefanalyzer~thus performs a two-level action identification, letting the LLM identify actions at each level rather than imposing a fixed decomposition.
First, it identifies low-level actions, such as dereferencing a pointer, which enables it to detect low-level vulnerabilities such as null-pointer dereferences.
Second, it identifies what the function aims to achieve as a whole, such as canceling a payment, which enables it to detect domain-specific vulnerabilities such as improper authorization.

\textbf{Belief Inference.}
For each identified action, \beliefanalyzer~infers the beliefs the developer assumes to be true when performing it.
A single action may provide evidence for multiple beliefs.
For example, as shown in Table~\ref{tab:belief-example}.right, the validation of the entity only against siteId and payment status (\emph{action}) embodies at least two distinct security assumptions (\emph{beliefs}): 
(\emph{B.1})~that the user who wants to cancel payment \texttt{paymentId} is sufficiently authenticated by the \texttt{siteId}, and/or 
(\emph{B.2})~that the \texttt{siteId} is a trusted, server-side value that is not under attacker control.
\beliefanalyzer~infers each of them, since any of those may underlie a potential security flaw.

\subsection{Belief Evaluation}
\label{sec:evaluation}
Having inferred a set of beliefs, how can we use them to identify potential vulnerabilities?
Recall that a belief is an assumption the developer holds while implementing the code.
A potential vulnerability can therefore arise in two ways. The assumption itself may be unreasonable, or the assumption may be reasonable yet not upheld by the implementation.
\beliefanalyzer~addresses these two cases with two complementary checks, a \textit{sanity check} and a \textit{violation check}.

\textbf{Sanity Check.}
The sanity check assesses whether a belief is reasonable, i.e., whether it aligns with standard security practice.
If a belief does not align, no implementation built on it can be secure, and a potential vulnerability follows regardless of how carefully the code is written.
Beyond a Reasonable or Unreasonable label, the sanity check also produces an explanation of which security practice the belief is compared against and why it is considered reasonable or not.
For the \texttt{cancel} function in Listing~\ref{lst:func-cancel}, the belief B.1 is unreasonable, because site-level and payment-level authorization are different. Note that vulnerabilities induced by such unreasonable beliefs are \emph{context-independent}: Any code that relies on them is vulnerable.

\textbf{Violation Check.}
Even when a belief is reasonable, the implementation may fail to uphold it under certain conditions, and a potential vulnerability can still arise.
Rather than simply judging whether such a condition exists, the violation check derives the specific conditions under which a belief no longer holds.
Since \beliefanalyzer~operates at the function level, a violation condition may be mitigated by upstream code, so \beliefanalyzer~surfaces these conditions explicitly for further verification.
For the \texttt{signInCredentials} function (Listing~\ref{lst:func-signin}), the developer is aware of user enumeration and follows the standard practice of returning a generic error.
However, the implementation overlooks a timing side-channel induced by an additional function call, under which the belief is violated and the vulnerability that was meant to be mitigated may still occur.
\beliefanalyzer~derives the violation conditions for every belief, including those already judged unreasonable, to fully characterize how each belief may fail.

\subsection{Vulnerability Reporting}
\label{sec:identification}

The sanity and violation checks together flag the beliefs that may lead to a potential vulnerability: a belief that is unreasonable, or one whose violation condition can be triggered, indicates a potential vulnerability.
Based on these flagged beliefs, \beliefanalyzer~produces a \textit{report of potential vulnerabilities}.
The report consists of a set of \textit{findings}, each describing one potential vulnerability together with an \textit{explanation} of its root cause and the context under which it can be exploited.
Since a single potential vulnerability may be indicated by multiple beliefs, each finding is linked back to the beliefs that support it, and overlapping beliefs are consolidated into one finding.
We call these vulnerabilities \textit{potential} because the analysis is performed at the function level, and whether a potential vulnerability is real depends on the surrounding context.

%% file: tables/Prompt.tex
\begin{figure}[t]
\begin{tcolorbox}[colback=white, colframe=black!60,
  boxrule=0.4pt, arc=1pt, left=4pt, right=4pt, top=3pt, bottom=3pt,
  fontupper=\footnotesize\ttfamily]
\textbf{Task:} Analyze the beliefs the developer holds when performing
actions in the code, analyze their reasonableness, identify their
violation conditions, and based on this analysis identify all
potential vulnerabilities.

\medskip
\textbf{Reasoning Steps:}\\
(1) \textbf{Action Identification:} Identify the operations performed
by the code.\hfill{\color{gray}\S\ref{sec:inference}}\\
(2) \textbf{Belief Inference:} For each action, infer the beliefs that
developers may hold.\hfill{\color{gray}\S\ref{sec:inference}}\\
(3) \textbf{Sanity Analysis:} Analyze the reasonableness of these
beliefs.\hfill{\color{gray}\S\ref{sec:evaluation}}\\
(4) \textbf{Violation-Condition Derivation:} Identify the violation
conditions for these beliefs.\hfill{\color{gray}\S\ref{sec:evaluation}}\\
(5) \textbf{Potential Vulnerability Identification:} Based on the
belief analysis, identify all potential security vulnerabilities in
the given code.\hfill{\color{gray}\S\ref{sec:identification}}

\medskip
\textbf{Definitions:}\\
-- Belief: A fact that the developer assumes to be true when
performing an action.\\
\ldots

\medskip
\textbf{Function:} \textit{\{\{function\_code\}\}}\\
\textbf{Output Format:} \textit{\{\{output\_format\}\}}
\end{tcolorbox}
\caption{Prompt of \textsc{Creed} (condensed). Verbatim prompt and
output schema are in the artifact.}
\label{lst:prompt}
\end{figure}




%% file: tables/FuncCancel.tex
\begin{lstlisting}[
  style=base,
  escapeinside={(*@}{@*)},
  linebackgroundcolor={%
  \ifnum\value{lstnumber}=3\color{hlline}%
  \else\ifnum\value{lstnumber}=5\color{hlline}%
  \else\color{codebg}%
  \fi\fi},
  caption={Improper authorization in PublicCMS (CVE-2026-2010)},
  label={lst:func-cancel}
]
public boolean cancel(short siteId, long paymentId) {
  TradePayment entity = getEntity(paymentId);
  if (null != entity && siteId == entity.getSiteId()
  && entity.getStatus() == STATUS_PENDING_PAY) {
    entity.setStatus(STATUS_CLOSE);
    ...
}
\end{lstlisting}

%% file: tables/BeliefExample.tex
\begin{table}[th]
\centering
\small
\setlength{\tabcolsep}{2.5pt}
\caption{\beliefanalyzer's analysis of two example functions, \texttt{signInCredentials()} and \texttt{cancel()}.}
\label{tab:belief-example}
\begin{tabularx}{\linewidth}{P{0.12\linewidth}P{0.4\linewidth}L}
\toprule
\textbf{Analysis} & \textbf{\texttt{signInCredentials()}} (CVE-2026-25222) & \textbf{\texttt{cancel()}} \newline (CVE-2026-2010) \\
\midrule
Action & Returns the same error for both missing-user and wrong-password cases. & Validates the entity only against siteId and payment status.  \\
\midrule
Belief & A generic error response makes the two failure modes indistinguishable to attackers. & A matching siteId is sufficient authorization to cancel this payment. \\
\midrule
Sanity & [Reasonable] a standard practice to prevent user enumeration. & [Unreasonable] site-level and payment-level authorization are different. \\
\midrule
Violation & Response time differs significantly between two failure modes. & A same-site user (matching siteId) cancels a payment they do not own. \\
\bottomrule
\end{tabularx}
\end{table}

%% file: sections/0x30Metric.tex
\section{Evaluating Function-Level Vulnerability Detection}
\label{sec:function-level-eval}

\subsection{Metrics for Function-Level Vulnerability Detection}\label{sec:metrics}

To evaluate function-level vulnerability detection ability, we do not adopt the classic metrics based on binary classification, as recent work has shown that such an evaluation is fundamentally invalid for evaluating machine-learning-based vulnerability detection~\cite{metric:niklasbench}. 
Specifically, whether a function causes a vulnerability in a program or not often depends on the context within which this function is called. For example, every caller of a given function may always make sure that a pointer points to allocated memory before it is dereferenced in the function (with a potential null-pointer-dereference vulnerability).
Without knowing this context, it is not possible to decide whether the function causes the program to be vulnerable or not.

However, our work addresses the context-dependency problem for function-level vulnerability detection using a \emph{context-conditional} approach that offers explanations for multiple findings rather than a single binary decision. \beliefanalyzer reports each \textit{finding} with an \textit{explanation} of the root cause and the conditions under which a function may cause a security issue.

\textbf{Matched finding.} During experimental evaluation, we have access to substantial evidence for the given vulnerability, including the CVE description, the commit message explaining the vulnerability and how it is fixed, and the patch itself. Using this evidence, we can determine the context condition and its root cause.
Since both the context condition and the root cause are explained by the finding, we can determine whether an auto-generated vulnerability finding is consistent with that evidence. We call such findings \textit{matched}. By checking both the condition and the root cause, the context-dependency problem is thus addressed. 

\begin{figure}[ht]
    \centering
    \includegraphics[width=.8\linewidth]{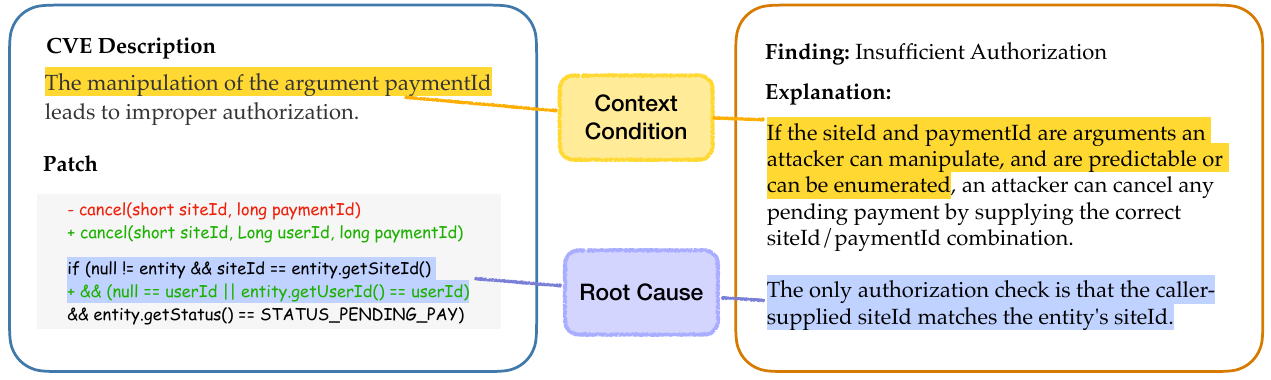}
    \caption{Matching a finding's explanation (right) to the known evidence (left).}
    \label{fig:cve-explanation}
\end{figure}

\textbf{Measuring effectiveness} (recall). We can measure the effectiveness of an explanation-based vulnerability detection approach using \emph{recall} which is defined as the proportion of known vulnerabilities for which the approach reports at least one matched finding.
Formally, the recall of an approach $A$ is defined as
\[
\mathrm{Recall}(A)
=
\frac{
{N}^A_{\mathrm{match}}
}{
{N}_{\mathrm{vuln}}
},
\]
where $N_{\mathrm{vuln}}$ is the total number of known vulnerabilities in the benchmark and $N^A_{\mathrm{match}}$ is the number of reports for which approach $A$ reports at least one matched finding.

Following PrimeVul~\cite{agent4static:primevul}, we also measure whether detections are specific to the defect by analyzing each vulnerable function together with its patched version, which differ only in the fix.
We adapt the binary paired evaluation to explanations: an approach is correct on a pair if it reports a matched finding on the vulnerable function and none on the patched one (\textbf{P-C}). The other outcomes are a match on both (\textbf{P-V}), on neither (\textbf{P-B}), or only on the patched function (\textbf{P-R}). A high P-C count indicates that the vulnerability is reported because of the defect, not other features of the function.

Recall remains our primary measure, as it directly captures whether an approach finds and explains known vulnerabilities, while the paired outcomes qualify it.
This allows us to evaluate how well an approach detects known vulnerabilities.
However, most approaches typically generate multiple findings per function.
How do we estimate the human effort required to identify a known vulnerability by going through all these findings?

\textbf{Measuring effort} (overhead). We can measure the effort of an explanation-based vulnerability detection approach using \emph{overhead} which is defined as the average number of findings to be examined before confirming one known vulnerability that we expect to find.
By assuming all unmatched findings are incorrect, this worst-case human effort can be captured by the overhead of an approach. 
Formally, the overhead of an approach $A$ is defined as
\[
\mathrm{Overhead}(A)
=
\frac{
{N}^A_{\mathrm{finding}}
}{
{N}^A_{\mathrm{match}}
},
\]
where $N^A_{\mathrm{finding}}$ is the total number of findings reported by $A$.

Overhead is a pessimistic upper bound, as it treats every unmatched finding as a false alarm. 
In practice, some unmatched findings may still be plausible vulnerabilities, so the actual inspection effort can be lower than this worst-case overhead (see RQ-2). 

\subsection{Evaluation Procedure}

\begin{figure}[ht]
    \centering
    \includegraphics[width=0.45\linewidth]{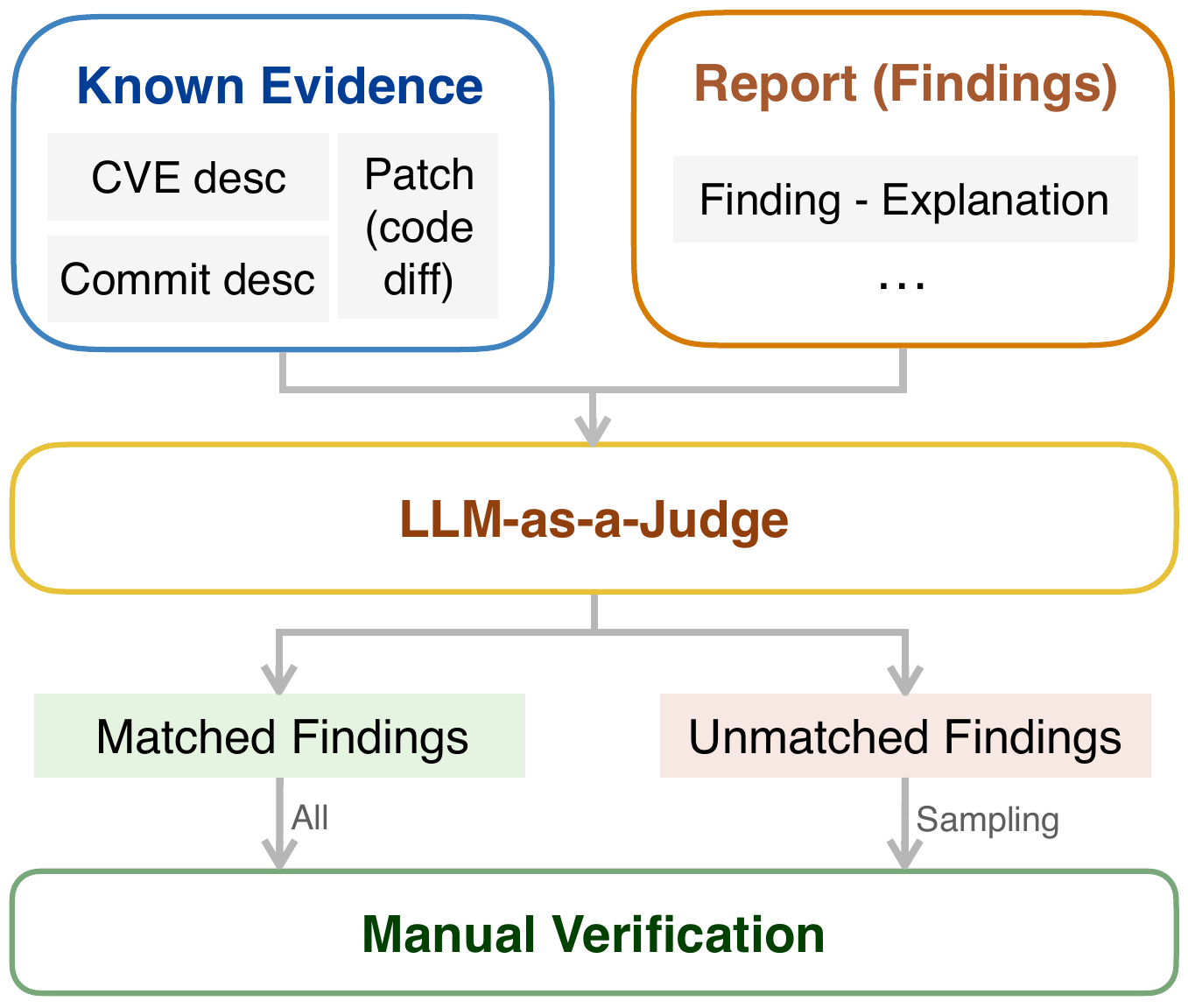}
    \caption{Overview of our evaluation procedure}
    \label{fig:eval-protocol}
\end{figure}

Figure~\ref{fig:eval-protocol} illustrates our evaluation procedure. 
We construct a benchmark from real CVEs, where each entry consists of a vulnerable function together with its known evidence: the CVE description, the commit message, and the patch. 
Each evaluated approach analyzes the function and produces a report, i.e., a list of findings, where each finding identifies a potential vulnerability with an explanation of the root cause and the condition under which it occurs.

\textbf{LLM-as-a-Judge.} Given such a report, how do we determine whether it explains the known vulnerability? Manually inspecting every report is expensive and hard to scale, so we employ an LLM-as-a-Judge~\cite{metric:llmjudge,metric:llmjudgese}. 
To avoid self-preference bias, where a model tends to favor outputs from its own family, we use a judge from a different model family than the evaluated approaches. 
The judge receives the known evidence and the report, and labels each finding as \textit{matched} if its explanation is consistent with the known evidence, and as unmatched otherwise.

\textbf{Manual check.}
To ensure the soundness of our results, we manually recheck the judge's outputs. For every finding the judge marks as matched, we manually verify the match, so that the reported recall is not inflated by erroneous judgments (see RQ-5).
For the unmatched findings, which are too numerous to inspect exhaustively, we sample a subset and manually assess whether each is plausible (see RQ-2).

%% file: sections/0x40Evaluation.tex
\section{Evaluation}

We evaluate \beliefanalyzer~by investigating how effectively it detects vulnerabilities, how much inspection effort its reports demand, across which vulnerability types it succeeds, and how each of its components contributes to this performance.

\begin{itemize}[leftmargin=10pt,itemsep=2pt]
    \item \textbf{RQ-1 (Effectiveness).} How effective is \beliefanalyzer~at detecting vulnerabilities compared to state-of-the-art LLM-based auditing approaches?
    \item \textbf{RQ-2 (Trade-off).} How does \beliefanalyzer~balance detection effectiveness against the effort of inspecting its reports?
    \item \textbf{RQ-3 (Vulnerability Types).} \begin{itemize}
    \item (a) How does the effectiveness of \beliefanalyzer and the other tools vary across different types of vulnerabilities?
    \item (b) How often can a vulnerability be expressed as a general pattern useful for traditional static analysis?
    \end{itemize}
    \item \textbf{RQ-4 (Component).} How does each component of \beliefanalyzer~contribute to its performance?
    \item \textbf{RQ-5 (Judge Reliability).} How reliable is the LLM-as-a-Judge at deciding whether a finding is matched?
\end{itemize}

\subsection{Evaluation Setup}

\textbf{Benchmark.}
To mitigate data leakage from the LLM training set, we construct our benchmark from recently disclosed vulnerabilities (as of the time of writing).
We use CVEfixes~\cite{dataset:cvefixes} to collect CVEs were disclosed between September 2025 and March 2026, yielding 2751 CVEs.
We then extract the vulnerable functions modified in their fixing commits.
Each instance consists of a vulnerable function, which serves as the input to each approach, together with its CVE description and fixing commit, which serve as the ground truth for evaluation.
To ensure that a collected function is genuinely the one fixed by the CVE, we retain only CVEs where (1) the function is the only one changed by the commit, or (2) the CVE description explicitly names the function, or names its file and the function is the only one changed in that file.
This yields 903 vulnerable functions spanning 622 CVEs.
Since these heuristics do not guarantee that a function contains the defect logic of its CVE, we randomly sample functions and manually confirm that each function contains the root cause of its CVE, until we reach 50 confirmed functions. In total, we analyzed 60 sampled functions, 10 (17\%) were discarded as merely co-modified or as callers of the defective function.

The final benchmark spans 45 distinct projects, 10 programming languages, and 10 vulnerability types.
Figure~\ref{fig:typerecall} shows the distribution of vulnerability types in our benchmark, of which the majority (40 of 50) are semantic vulnerabilities rather than memory-related ones. 
Many of these are also beyond what a general static-analysis pattern can express, which we analyze in detail in RQ-3 (Section~\ref{sec:rq3}).
The distribution closely tracks that of the 622-CVE pool (percentages in parentheses in Figure~\ref{fig:typerecall}), suggesting that the 50 functions are representative of recent CVEs.\footnote{Pool types are mapped from NVD-assigned CWEs. On the benchmark, this mapping agrees with our manual labels in 96\% of cases.}


\textbf{Tools and Techniques.}
We compare \beliefanalyzer~with four baselines that span different paradigms of LLM-based vulnerability detection, none of which implement a belief-guided analysis.
\begin{itemize}[leftmargin=0.34cm,itemsep=3pt]
\item \emph{RepoAudit}~\cite{llm4static:repoaudit} represents the syntactic static-analysis paradigm, an LLM-assisted repository-level auditing tool that follows a traditional static-analysis workflow.
\item \emph{\primevulexp}~\cite{agent4static:primevul} represents prompt-based chain-of-thought reasoning
, where the LLM is declared an expert in static program analysis and asked to provide, if any, the vulnerabilities, reasons, as well as the context and finally a binary label (i.e., vulnerable or not). To give this baseline a fair advantage, we do not evaluate the label (as in the original PrimeVul-CoT work) but the entire explanation as our evaluation requires.
\item \emph{LLM-Thinking} enables the LLM's thinking mode~\cite{claudethinking} and features a prompt similar to \primevulexp to find vulnerabilities like a security expert. 
\item \emph{VulTrial}~\cite{agent4static:vultrial} represents the multi-agent paradigm, a framework that audits code through multiple cooperating agents. One agent works like \emph{\primevulexp} and reports potential vulnerabilities while the three other agents debate and filter these reports.
\end{itemize}
For \emph{VulTrial}, \emph{\primevulexp}, and \emph{LLM-Thinking} what exactly constitutes a vulnerability remains undefined while for \beliefanalyzer a vulnerability is defined as an unreasonable or breakable belief.

\textbf{Model and Settings.}
For all baselines, we use Claude Sonnet 4.6 (model ID claude-sonnet-4-6) as the backbone LLM so that performance differences can be explained by the approaches rather than the underlying model.
The model's thinking mode is enabled only for LLM-Thinking, with medium effort.
We set the temperature to 0.0 to reduce randomness~\cite{llm4static:repoaudit, agent4static:vultrial, prompt:chatgpt}, except for LLM-Thinking, for which the temperature is not configurable in thinking mode~\cite{claudethinking}.
As the model for our \emph{LLM-as-a-judge}, we use GPT-5.4 (model ID gpt-5.4-2026-03-05), which is from a different model family than the evaluated approaches.

\textbf{RepoAudit Configuration}. To support RepoAudit's repository-level design, we provide it with the full repository for context.
It also relies on user-specified source functions, which we manually identify and provide where applicable. 


\subsection*{RQ-1: Effectiveness}

\input{tables/ResultMain}

\textbf{Results.} Table~\ref{tab:main-results}.left shows \beliefanalyzer's effectiveness in terms of recall compared to the four baselines. \beliefanalyzer~achieves the highest recall, successfully identifying and correctly explaining the root cause of 39 (78\%) of the 50 vulnerabilities. 
On the paired evaluation, it also reaches the highest number reports for vulnerable functions that disappear on the patched counterparts (33 of 50 P-C), confirming that the reports are caused by the defect.
Since all approaches share the same backbone model (Claude Sonnet 4.6), this improvement indeed reflects the effectiveness of our belief-guided reasoning approach rather than a difference in the strength of the model.

In comparison, \emph{RepoAudit} follows the (syntactic) static-analysis paradigm and supports only memory-related vulnerabilities in a subset of the benchmark's languages.\footnote{Only 28 of the 50 functions are in languages RepoAudit supports, and only two of them (one use-after-free and one null-pointer dereference) fall within the memory-related types its checkers target.}
It is therefore not capable of addressing the semantic vulnerabilities that dominate the benchmark.
While \emph{VulTrial}, the multi-agent framework, is language-agnostic and indeed capable of finding semantic vulnerabilities, it still \emph{fails} to identify the expected vulnerability in the majority of cases.
\primevulexp~and LLM-Thinking reach higher recall, but they have no instructions that define what constitutes a vulnerability. So their findings focus on common issues and overlook root causes that depend on specific code context and implementation.

\textbf{Explanation.} The advantage of \beliefanalyzer~comes from its belief-guided reasoning. Take the \texttt{signInCredentials} function as an example (Listing~\ref{lst:func-signin}; Table~\ref{tab:belief-example}.left). The other techniques identify vulnerabilities generically and report a range of common authentication weaknesses, but none uncovers the timing-based root cause. \beliefanalyzer~does not stop at recognizing the generic error as a mitigation of a potential user enumeration vulnerability. It infers the belief behind it (a generic error makes the two failure cases indistinguishable), and explicitly derives when this belief fails. It further reveals that the error messages are identical but the response times are not, so user enumeration remains possible.

\result{\textbf{RQ-1}. \beliefanalyzer~achieves the highest recall, correctly identifying and explaining 39 of 50 root causes (78\%), outperforming the strongest baseline by 10 percentage points.}

\subsection*{RQ-2: Effort and the Effectiveness Trade-off}

\textbf{Results (Effort)}. Table~\ref{tab:main-results}.right shows the effort required to go through \emph{all} vulnerability findings for each function. 
The overhead and total number of findings show that to identify all 39 vulnerabilities which \beliefanalyzer can find, a security researcher would have to go through at most 310 findings,\footnote{We note that 8 of the 310 findings were duplicates.} i.e., about 8 findings for each of the 39 discovered vulnerabilities \emph{that we expect to find}. Compared to LLM-Thinking, this is a reduction in overhead of about 30\%.
We note that due to an unmatched finding may still describe \emph{an unexpected vulnerability}, we cannot treat all of them as false positives. We estimate their true share in the precision analysis below.
On the machine side, \beliefanalyzer~needs a single LLM call and about 8.5k tokens per function, three times \primevulexp~and twice LLM-Thinking, but less than half of the multi-step approaches, which make four to five calls per function.

In comparison, \emph{VulTrial} has the least overhead with 3.3 findings for the security researcher to go through for each of the 21 discovered and expected vulnerabilities. This is due to the substantial additional effort by the three other agents in the multi-agent setup to debate and filter the findings reported by the first agent. 
To evaluate how the multi-agent approach would reduce the overhead for our belief-guided approach, we evaluated \emph{VulTrial} where the security agent was replaced by \beliefanalyzer. Indeed, we find that the overhead for \emph{VulTrial(\beliefanalyzer)} reduced to 2.8 findings per discovered vulnerability. However, recall also reduced to 0.36 (18/50).

Approaches with higher recall also tend to come with higher overhead, while the approach with the lowest overhead misses almost half of the vulnerabilities that the approach with the best recall (\beliefanalyzer).
In the following, we analyze the recall-overhead trade-off: Which approach finds the most vulnerabilities with the least inspection effort?

\begin{figure}[ht]
\begin{minipage}[t]{0.55\linewidth}
  \centering
  \includegraphics[width=\linewidth]{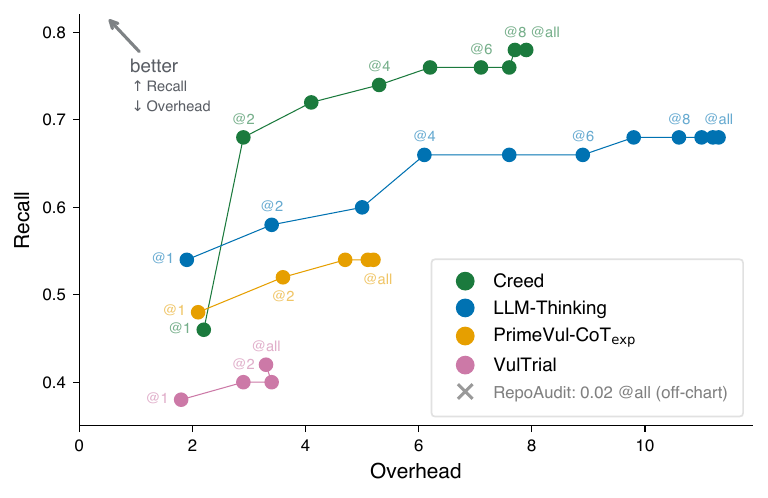}
  \caption{Recall and overhead when only the top-$k$ findings per function are inspected (@$k$). The upper-left corner is better.}
  \label{fig:topN}
\end{minipage}\hfill
\begin{minipage}[b]{0.43\linewidth}
  \input{tables/ResultTopk}

\end{minipage}
\end{figure}

\textbf{Presentation (Trade-off)}. 
Figure~\ref{fig:topN} visualizes the trade-off between recall and overhead as we consider the top-$k$ reports for every function for various values of $k$. Higher recall and lower overhead (top-left) are better.
None of the approaches has a ranking component. This first-$k$ selection simply follows the common inspection order, in which a human reads the findings sequentially~\cite{huamanauditor:sequentialread}.

\textbf{Results (Trade-off)}. Under a limited inspection budget, \beliefanalyzer~recovers the most vulnerabilities. If a security researcher were to look through at most two findings per function (@2), they would attain the \emph{highest recall} (0.68; 10 percentage points above the second-best) at an overhead of 2.9 for every discovered, expected vulnerability, placing it in the top-left-most corner of Figure~\ref{fig:topN}. Increasing the inspection budget to $k=3$ (@3), the recall of \beliefanalyzer increases by four percentage points to 0.72 and already exceeds what any other technique can achieve even if the security researcher was ready to spend unlimited effort.

\textbf{Results (Precision@1).} 
Since an unmatched finding may still describe \emph{an unexpected but actual vulnerability}, we manually inspect the first finding (@1) for every function of \beliefanalyzer which a human would read first when inspecting findings sequentially~\cite{huamanauditor:sequentialread}.
Based on this inspection, precision@1 is 0.74 (37/50), comprising 23 matched findings and 14 manually confirmed \emph{unexpected vulnerabilities}.
Eleven (11) of 14 were patched, and the remaining three we are preparing to responsibly disclose.
The other 13 are false positives of two kinds: ten are \emph{inter-procedural}, where \beliefanalyzer~derives the violation conditions correctly but the issue is mitigated by the caller or a called guard API; the other three are \emph{intra-procedural}, where \beliefanalyzer~overlooks an in-function constraint that prevents the risk. 
This suggests that the reported overhead is conservative, as many unmatched findings are real defects.


\result{\textbf{RQ-2}. 
\beliefanalyzer~offers the best recall-overhead trade-off. Inspecting only three findings per function, it exceeds what any other technique reaches with unlimited effort.}

\subsection*{RQ-3.a: Effectiveness across Vulnerability Types}
\label{sec:rq3}

\begin{figure}[ht]
    \centering
    \includegraphics[width=.7\linewidth]{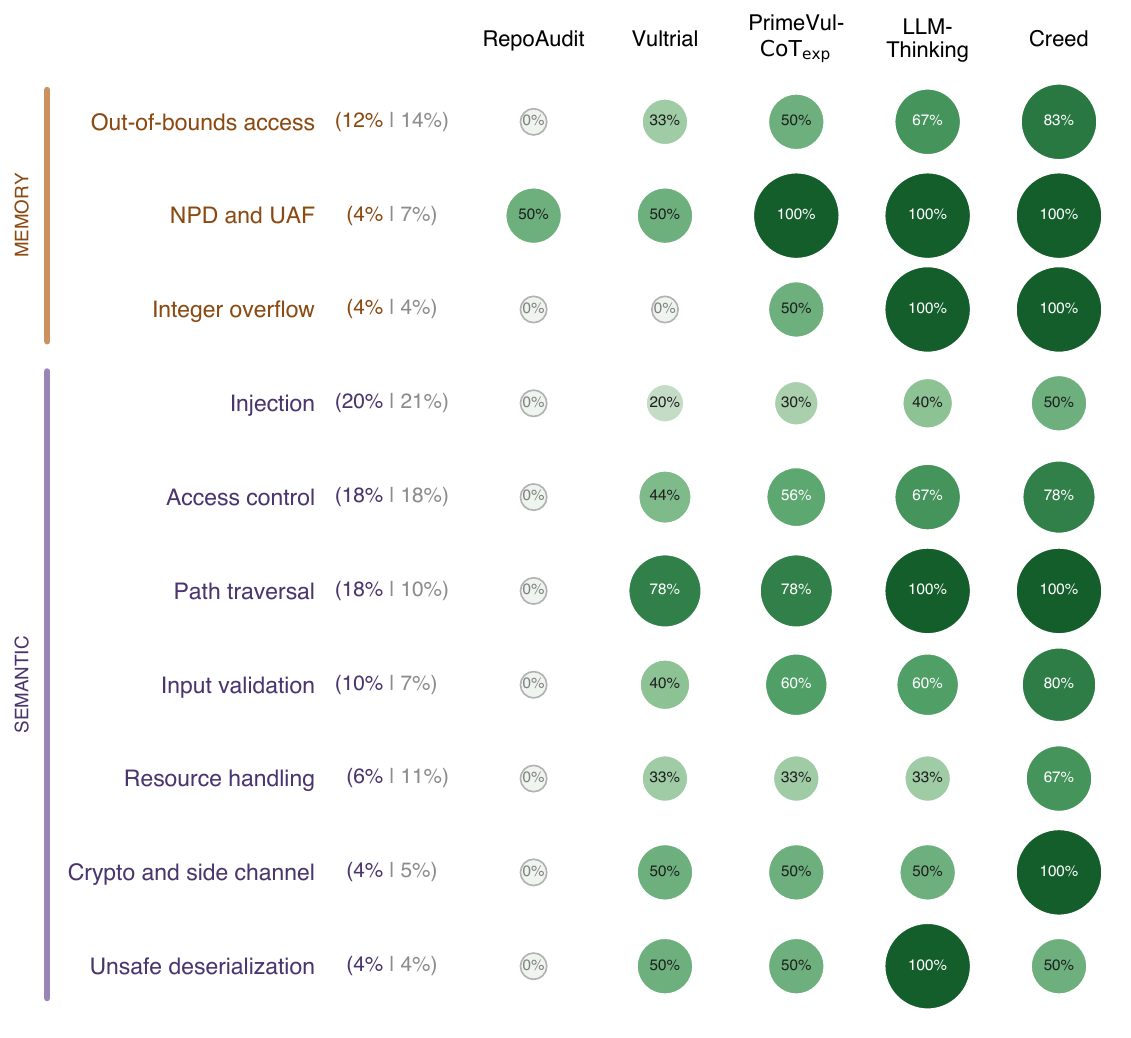}
    \caption{Per-type recall of each approach. In parentheses: share of the type in the benchmark (colored) and in the 622-CVE pool it was sampled from (gray). Rows are grouped into memory and semantic vulnerabilities.}
    \label{fig:typerecall}
\end{figure}

Figure~\ref{fig:typerecall} shows the distribution of vulnerability types within our vulnerability dataset, and the recall for each vulnerability type.
\beliefanalyzer~has the highest recall on almost every type, and it is the only approach that is strong on both the memory-related and semantic vulnerability types. It detects 90\% of the 10 memory-related and 75\% of the 40 semantic vulnerabilities, which is the best in each. It improves on the strongest baseline \emph{LLM-thinking} in all cases, except where the baseline already detects all vulnerabilities for that type.

RepoAudit succeeds only when a case fits one of its predefined vulnerability patterns. VulTrial, \primevulexp, and LLM-Thinking can cover both memory-related and semantic vulnerabilities.
\primevulexp~and LLM-Thinking perform well on types with visible code patterns, such as memory issues and Path Traversal. VulTrial is less consistent on memory-related cases, likely due to filtering. These three techniques have lower recall on Input Validation, Access Control, and Resource Handling, where the root cause depends on the meaning of a specific check, permission, or resource.

\result{\textbf{RQ-3.a}. \beliefanalyzer~leads on both memory-related and semantic vulnerabilities, and is the only approach to detect the timing side channel.}

\subsection*{RQ-3.b: Expressibility across Vulnerability Types}

Traditional static analysis detects bugs by recognizing corresponding patterns on structural representations of the code, such as the abstract syntax tree (AST), the control-flow graph (CFG), or the program dependence graph (PDG). How often can those vulnerabilities in our dataset be expressed as general patterns or rules?

\textbf{Presentation}. For each vulnerability, we ask whether there exists a general pattern that could facilitate detection by an idealized static analysis. 
We call a vulnerability \textit{expressible} if a general pattern exists. An expressible vulnerability can be detected by traditional static analysis using a general pattern. For instance, a use-after-free vulnerability can be detected if there exists a path in the PDG such that there is first a call to \texttt{free} on a pointer-type variable that is later used without being redefined in-between. An injection vulnerability can be detected in the PDG if there exists an information flow (i.e., a transitive control- or data-dependency) between a public source and a sensitive sink. We call an expressible vulnerability that requires user-provided annotations of the source and sink, like in the latter example, as \emph{custom}. Otherwise, we call an expressible vulnerability as \emph{generic}, like in case of the use-after-free.
Cases that are \emph{not expressible} depend on the application's authorization policy, the meaning of a value or resource, or the adequacy of a check.
Specifically, expressible vulnerability types cover all memory-related vulnerabilities, as well as information-flow-based semantic cases, such as classic injection or simple path traversal.

\textbf{Results (expressibility)}. Table~\ref{tab:typebreakdown} shows the results. Of the 50 cases, 29 are expressible, of which seven (7) need a user-provided annotation to identify a source or sink. For the remaining 21 vulnerabilities, the traditional, syntactic static-analysis paradigm falls short. For 42\% of the benchmark, there exists no general pattern that would help us find them using a traditional static analysis. Even among those 29 that are expressible, a quarter need a user-provided annotation of the source or sink, which cannot be assumed to exist.

\input{tables/ResultPattern}

\textbf{Results (language-specificity)}. In our evaluation, we assume an idealized static analyzer. In practice, a real analyzer faces further limits, since it must support the relevant languages and implement a checker for each pattern. Our static-analysis baseline, RepoAudit, illustrates this. Among the 29 expressible vulnerabilities it recovers only one (1). RepoAudit supports just three memory defects across four languages, so only two (2) cases (one use-after-free, one null-pointer-dereference) are in scope. Of those, it catches the use-after-free (after we supply a user-specified source function) and misses the null-pointer-dereference, where the null pointer's indirect use lies beyond its dataflow analysis.

\beliefanalyzer~has the highest recall for both groups of vulnerabilities: expressible and non-expressible. It finds 24 of the 29 expressible and 15 of the 21 non-expressible patterns, more than any of the other techniques. This dual strength follows from its design, whose two-level action identification covers both ends. Identifying low-level actions, such as how a value is used, lets \beliefanalyzer~handle the low-level expressible bugs, including most memory-related vulnerabilities. 
Identifying the whole-function action, together with belief inference and evaluation, lets it reason about the remaining non-expressible cases. Such cases hinge on whether the developer's assumption is reasonable and whether it holds, which is exactly what \beliefanalyzer's sanity and violation checks evaluate. 
The timing side-channel of CVE-2026-25222 (Listing~\ref{lst:func-signin}) is one such case, and only \beliefanalyzer~detects it.

\result{\textbf{RQ-3.b}. 42\% of the benchmark vulnerabilities are not expressible as general static-analysis patterns. \beliefanalyzer~performs best on both expressible and non-expressible cases.}

\subsection*{RQ-4: Component contribution}

\textbf{Ablation}. To study how each component of \beliefanalyzer~contributes to its performance, we in turn remove each of its three components from the prompt (Fig.~\ref{lst:prompt}): action identification\footnote{Without action identification, \beliefanalyzer~infers beliefs directly from the function code rather than from explicitly identified actions.}, the sanity check and the violation check. 

Table~\ref{tab:ablation} shows all components are necessary: Removing any component lowers recall and raises overhead. 
Action identification matters the most: without it, the model infers beliefs directly from the code, and recall drops the most (0.78 to 0.60) while the report contains the fewest findings (269), since the actions are what ground the inferred beliefs. The sanity check and the violation check examine a belief from two angles, whether it is reasonable and under what conditions it breaks; removing either drops recall to 0.66, as each catches vulnerabilities the other misses.

\input{tables/ResultAblation}

\textbf{Component output analysis}. To better understand the intermediate quality of \beliefanalyzer, we manually examine the output of each component.
For each of the 39 matched cases, we take their cited belief and 
check two properties of each component: i) its output is grounded in the output of the previous component, and ii) its output contributes to the matched finding.
For the 11 missed cases, we identify the first component whose output fails to lead to the known vulnerability.

Table~\ref{tab:ablation}.right shows the results. For the 39 matched cases, every component's output holds in 36. The three exceptions are two beliefs attached to a different action than the code they are grounded in, and one sanity judgment based on an assumption the belief does not state.
For the 11 missed cases, the failure lies mostly in action identification and violation-condition derivation.
In three cases \beliefanalyzer~fails to extract the relevant \emph{action}, so no belief is inferred or evaluated about it. 
Three cases arise when identifying an exploitable \emph{violation condition} requires application-specific expertise.
For CVE-2026-32761, recognizing the risk requires knowing that a user without download permission should not share a file, a rule of the application rather than of general security practice. 
Belief inference and the sanity check account for one missed case each. 
The remaining three occur after these components. For CVE-2026-23829, every component's output is correct, but the finding is dropped from the final report, and two are misjudgments by the LLM-as-a-Judge, whose reliability we evaluate in RQ-5.

\result{\textbf{RQ-4}. Only if all components in \beliefanalyzer are enabled do we get the highest recall and the lowest overhead. And each component's output holds in almost all matched cases.}

\subsection*{RQ-5: Reliability of LLM-as-a-Judge}
\label{sec:rq5}

Since our results above rely on the LLM-as-a-Judge to decide whether a finding matches the known vulnerability evidence, we assess how reliably its judgment aligns with human experts.
Two annotators (both PhD students in security) independently and blindly annotated a random sample of 50 findings drawn from all findings reported across the five methods, without access to the judge's decisions.

Both annotators reached complete agreement (100\%). Against this human ground truth, the judge agreed on 48 of 50 findings (96\%). The two disagreements (Creed, CVE-2025-11372; PrimeVul, CVE-2025-66563) were both findings that the annotators marked as matched but the judge rejected. 

Beyond this sample, we verified every finding the judge accepted as matched across the full evaluation, and found only one false acceptance (\primevulexp, CVE-2026-33985).

\result{\textbf{RQ-5.} The LLM-as-a-Judge reliably aligns with human expert judgment, agreeing on 96\% of sampled findings.}

%% file: tables/ResultMain.tex
\begin{table}[th]
\centering
\begin{threeparttable}
\caption{Effectiveness on the 50 pairs of vuln./patched functions, and effort per function (Metrics in \S\ref{sec:metrics}).}
\label{tab:main-results}
\small
\setlength{\tabcolsep}{4pt}
\begin{tabular}{@{}l@{}r@{ \ }r|r@{ }r@{ }r@{ }r|r@{ }r@{ }|r@{ }r@{}}
\toprule
\multirow{2.4}{*}{Method}
& \multicolumn{6}{c|}{Effectiveness}
& \multicolumn{4}{c}{Effort} \\
\cmidrule(lr){2-7}\cmidrule(lr){8-11}
& \textbf{Matched} & \textbf{Recall} $\uparrow$
& \textbf{P-C} $\uparrow$ & \textbf{P-V} $\downarrow$ & \textbf{P-B} $\downarrow$ & \textbf{P-R} $\downarrow$
& \textbf{\#Findings} & \textbf{Overhead} $\downarrow$ & \textbf{\#Calls} & \textbf{Tokens} \\
\midrule
RepoAudit        & 1  & 0.02 & 1 & 0 & 48 & 1 & 8 & 8.0  & 5.2 & 54.6k \\
VulTrial         & 21 & 0.42 & 20 & 1 & 29 & \textbf{0} & 69 & \textbf{3.3} & 4.0 & 17.3k \\
\primevulexp     & 27 & 0.54 & 27 & \textbf{0} & 23 & \textbf{0} & 141 & 5.2  & 1.0 & \textbf{2.7k} \\
LLM-Thinking     & 34 & 0.68 & 26 & 8 & 14 & 2 & 384 & 11.3 & 1.0 & 3.9k \\
\beliefanalyzer  & \textbf{39} & \textbf{0.78} & \textbf{33} & 6 & \textbf{11} & \textbf{0} & 310 & 7.9 & 1.0 & 8.5k \\
\bottomrule
\end{tabular}
\begin{tablenotes}\footnotesize
\item \#Calls and Tokens are per-function averages; Tokens include input, output, and thinking tokens.
\end{tablenotes}
\end{threeparttable}
\end{table}

%% file: tables/ResultTopk.tex
\centering\footnotesize
\setlength{\tabcolsep}{3pt}
\captionof{table}{Recall (R) and overhead (O) when only the first $k$ findings of each report are inspected.}
\begin{tabularx}{\linewidth}{lCCCCCC}
\toprule
\multirow{2}{*}{Method} & \multicolumn{2}{c}{@1} & \multicolumn{2}{c}{@2} & \multicolumn{2}{c}{@3} \\
\cmidrule(lr){2-3}\cmidrule(lr){4-5}\cmidrule(lr){6-7}
 & R $\uparrow$ & O $\downarrow$ & R $\uparrow$ & O $\downarrow$ & R $\uparrow$ & O $\downarrow$ \\
\midrule
RepoAudit       & 0.02 & 8.0 & -- & -- & -- & -- \\
VulTrial        & 0.38 & \textbf{1.8} & 0.40 & \textbf{2.9} & 0.40 & \textbf{3.4} \\
\primevulexp    & 0.48 & 2.1 & 0.52 & 3.6 & 0.54 & 4.7 \\
LLM-Thinking    & \textbf{0.54} & 1.9 & 0.58 & 3.5 & 0.60 & 5.0 \\
\beliefanalyzer & 0.46 & 2.2 & \textbf{0.68} & \textbf{2.9} & \textbf{0.72} & 4.1 \\
\bottomrule
\end{tabularx}
\label{tab:ranking}


%% file: tables/ResultPattern.tex
\begin{table}[th]
\centering
\caption{Detected Vulnerabilities by Pattern-Expressibility}
\label{tab:typebreakdown}
\small
\setlength{\tabcolsep}{4pt}
\begin{tabular}{lrrr}
\toprule
\multirow{2.4}{*}{Method}
& \multicolumn{2}{c}{Expressible} & \multirow{1}{*}{Not Expressible} \\
\cmidrule(lr){2-3}
& Generic (22) & Custom (7) & \multicolumn{1}{c}{(21)} \\
\midrule
RepoAudit        & 1  & 0          & 0 \\
VulTrial         & 8  & 4          & 9 \\
\primevulexp     & 14 & 4          & 9 \\
LLM-Thinking     & 16 & \textbf{6} & 12 \\
\beliefanalyzer  & \textbf{18} & \textbf{6} & \textbf{15} \\
\bottomrule
\end{tabular}
\end{table}

%% file: tables/ResultAblation.tex
\begin{table}[ht]
\centering
\begin{threeparttable}
\caption{Ablation and component output analysis. Left: results without the step. Right: on matched cases, whether the step's result is correct; on missed cases, the first step that fails.}
\label{tab:ablation}
\small
\setlength{\tabcolsep}{4pt}
\begin{tabular}{lcccc|cc}
\toprule
\multirow{2}{*}{\textbf{Step}}
& \multicolumn{4}{c|}{Ablation (w/o step)} & \multicolumn{2}{c}{Component output analysis} \\
& \textbf{Matched} & \textbf{Recall} $\uparrow$ & \textbf{\#Findings} & \textbf{Overhead} $\downarrow$
& \textbf{Holds} & \textbf{Fails} \\
\midrule
Action    & 30 & 0.60 & 269 & 9.0 & 39 & 3 \\
Belief    & -- & --   & --  & --  & 37 & 1 \\
Sanity    & 33 & 0.66 & 304 & 9.2 & 38 & 1 \\
Violation & 33 & 0.66 & 312 & 9.5 & 39 & 3 \\
\midrule
Full \beliefanalyzer & \textbf{39} & \textbf{0.78} & 310 & \textbf{7.9} & 39 & 11\tnote{*} \\
\bottomrule
\end{tabular}
\begin{tablenotes}\footnotesize
\item[*] Three misses occur after these components: one is dropped at consolidation, two are misjudged.
\end{tablenotes}
\end{threeparttable}
\end{table}


%% file: sections/0x60RelatedWork.tex
\section{Related Work}

\subsection{Belief-Guided Bug Detection}
Inferring developer beliefs from code to find bugs has been explored for more than two decades in static analysis.
Engler et al.~\cite{static:deviant} pioneered this view by extracting beliefs with manually designed templates and reporting statistical deviations as potential bugs.
Later work extends this idea beyond hand-written templates by using automatic inference techniques to extract developer beliefs from code~\cite{belief:specification,belief:prminer,belief:simcode}.
This idea has also been specialized to concrete bug domains, such as beliefs about API post-handling in API misuse~\cite{belief:APHP}, error-return conventions and recovery actions in error handling~\cite{belief:errorhandling}, as well as required validation or security checks~\cite{belief:crix,belief:ippo}.
Although these works use different signals and target different bugs, they share the same pattern-based, syntactic view.
They first turn developer beliefs into patterns, and then find bugs by checking deviations of them. We evaluate the prospects of this approach in RQ-3.b and discuss it further in Section~\ref{sec:discussion}.

\subsection{LLM-assisted Static Bug Detection}
At a high level, static bug detection typically proceeds in three steps: identifying bug-relevant program entities, constructing the analysis context around these entities, and applying bug detection rules to report bug candidates.
Recent LLM-based approaches augment different stages of this workflow.
One line of work uses LLMs for \textit{bug-relevant entity identification}, leveraging their semantic understanding to locate suspicious APIs~\cite{llm4static:inferroi}, sources and sinks~\cite{llm4static:iris}, or key variables and statements~\cite{llm4static:gptscan}. Another line uses LLMs to assist \textit{analysis-context construction}, for example, by generating graph queries for code property graphs~\cite{llm4static:llmxcpg} or guiding program slicing~\cite{llm4static:protocolguard}. A third line uses LLMs for \textit{detection-rule construction}. For example, KNighter~\cite{llm4static:knighter} synthesizes static analysis checkers from historical bug-fixing patches, extending pre-defined human-written patterns with generated patterns. Finally, more recent end-to-end systems further employ LLMs to orchestrate the entire workflow~\cite{llm4static:llift,llm4static:latte,llm4static:repoaudit}.
Although these works reduce manual effort, they assume that bugs can be sufficiently captured by patterns. However, vulnerabilities often arise from semantic flaws, which patterns cannot completely capture.

\subsection{Direct LLM-based Vulnerability Auditing}

\textbf{Prompt engineering.}
Recent works apply general prompt engineering techniques to vulnerability detection, including CoT for step-by-step vulnerability reasoning~\cite{prompt:vsp, agent4static:primevul}, tree-of-thoughts for exploring multiple reasoning paths~\cite{prompt:totselfconsistency}, and in-context learning for demonstration-based guidance~\cite{prompt:chatgpt}.
Beyond these general techniques, some works tailor prompts with code-specific evidence, such as data flow, control flow or API call sequence~\cite{llm4static:prompt}.
While such prompt designs can improve LLMs' reasoning, they mainly elicit generic reasoning steps rather than bug-detection-specific reasoning.

\textbf{Role-based agents.}
Multi-agent approaches decompose the auditing process into different roles: one agent identifies potential vulnerabilities, while others critique or validate the findings. GPTLens~\cite{agent4static:gptlens} employs a collaboration between an auditor and a critic, and VulTrial~\cite{agent4static:vultrial} instantiates roles such as auditor, defender, and judge. 
Similar role specializations have been explored in specific fields, such as smart-contract analysis~\cite{agent:iaudit,agent:llmsmartaudit}.
A role specialization can better elicit LLMs' semantic reasoning and improve the validation of candidates identified by the auditor agent. However, the initial auditing step still largely depends on a single agent's heuristic reasoning. As a result, subtle or novel bugs may be missed before the debate or critique stage begins. We confirmed this finding in RQ-1 when we replaced the vulnerability-reporting agent in VulTrial with our \beliefanalyzer.

\textbf{Knowledge-enhanced LLMs.}
Another line of work enhances LLMs with additional vulnerability knowledge.
Fine-tuning-based approaches update the model's parameters on labeled vulnerability data. VulLLM~\cite{finetune:vulllm} adds vulnerability localization and interpretation as auxiliary tasks to capture root causes, while R2Vul~\cite{finetune:r2vul} distills structured, security-aware reasoning into small models via reinforcement learning.
Such adaptation can improve accuracy, but fine-tuned models often learn surface patterns rather than real security logic, and generalize poorly to new code.
RAG-based approaches augment LLMs with external vulnerability knowledge at detection time.
Vul-RAG~\cite{agent4static:vulrag} retrieves relevant knowledge from historical cases, while VulInstruct~\cite{agent4static:vulinstruct} extracts security specifications from past vulnerabilities and retrieves them during detection.
Such retrieval can supply domain-specific knowledge to the model, although irrelevant or incomplete retrieval can still bias it toward incorrect judgments.

\textbf{Our belief-guided reasoning.}
\beliefanalyzer~takes a different view of how LLMs should be used for vulnerability auditing. 
Instead of using LLMs as components in a static-analysis pipeline or directly prompting them to generate vulnerability reports, \beliefanalyzer~reframes vulnerability auditing as the discovery of unreasonable or incorrect developer beliefs, guiding LLMs to reason about why code is buggy rather than whether it matches a known pattern.

%% file: sections/0x50Discussion.tex
\section{Discussion}
\label{sec:discussion}

\subsection{Beyond Rule-Based Detection}

We consider static analysis as the analysis of source code without executing it, and are convinced that our belief-guided static analysis offers new opportunities for identifying semantic bugs that have previously been out of reach.

In the security domain, our randomly sampled benchmark shows that most of the vulnerabilities are semantic (40 of 50), and among them over 50\% (21/40) are difficult to express as a rule. 
This means that a large proportion of domain-specific vulnerabilities cannot be captured by the traditional static analysis paradigm based on patterns or rules, since they depend on the specific context (such as the timing side-channel that is only vulnerable in the sign-in authentication context). 
With the rule-based, language-specific view these domain-specific contexts are reduced or hard to recognize, and thus the vulnerabilities that need more semantic meaning are difficult to detect.

\beliefanalyzer~addresses this problem through belief-guided vulnerability detection. Rather than fixing rules before detection, it derives the violation conditions while evaluating each belief. Since each belief is inferred from an action and grounded in the specific context, \beliefanalyzer~can identify semantic vulnerabilities such as the timing side-channel and the insufficient permission check that rule-based detection overlooks (see Table~\ref{tab:belief-example}). 

In this paper, we explore the opportunities of LLMs for program analysis and find that they are indispensable when it comes to semantic, high-level, domain-specific, language-agnostic reasoning about developer intention and beliefs.

\subsection{The Boundary of Belief-Guided Vulnerability Detection}

The main boundary of belief-guided vulnerability detection is that evaluating a belief sometimes requires knowledge that lies beyond common security practice and cannot be inferred from the code alone.
This appears in both the sanity check and the violation check.
For the sanity check, an application may have its own specified security policy or threat model beyond common security practice, making it hard to judge whether a belief is reasonable.
For \texttt{withPermShare} (CVE-2026-32761), judging that checking only the share permission is insufficient requires the application's specified policy that a user without download permission should not share a file.
For the violation check, a belief may be violated under open-ended conditions, and knowing which of them can actually occur often requires context beyond the function.
For \texttt{containsDangerousSubstitution} (CVE-2025-65946), the dangerous substitutions are essentially unbounded, so \beliefanalyzer~lists many conditions but misses the one that applies.
Such knowledge lies outside the code. Improving belief-guided vulnerability detection therefore depends on providing the application's specified security policy and trust model.


\subsection{Threats to Validity}

\textbf{Construct validity}.
In our study, we evaluate each approach with two metrics, recall and overhead, both of which we report conservatively.
For recall, a report counts as matched only if a human confirms it, so the reported recall is a lower bound on the true recall.
For overhead, we treat every unmatched finding as a false alarm, so the reported overhead is a worst-case upper bound and the actual inspection effort is no higher (Section~\ref{sec:function-level-eval}).
Deciding whether a finding matches the known evidence relies on an LLM-as-a-Judge~\cite{metric:llmjudge,metric:llmjudgese}. Like for any LLM-based evaluation, there is inherent uncertainty: the LLM may misjudge a match.
To mitigate this, a) we choose the judge from a different model family than the evaluated approaches, and b) we manually verify the matches and assess the judge's reliability against human experts in RQ-5.

\textbf{Internal validity}.
A key concern is whether our experimental setup compares \beliefanalyzer~and the baselines fairly.
The first such factor is the underlying model. To ensure that the comparison reflects the approaches rather than the model, all approaches use the same backbone model. A robustness experiment with a different backbone (DeepSeek V4 Pro) shows that \beliefanalyzer~still achieves the highest recall, suggesting that the result is not specific to a single model.
The second is the non-determinism inherent in LLMs, which could make the results unstable. We mitigate this by setting the temperature to 0 for all approaches, except for LLM-Thinking, for which the temperature is not configurable in thinking mode.
A third factor is the differing settings some baselines require. RepoAudit performs repository-level analysis, so we provide it with the full repository and manually supply the user-specified functions it needs. 
PrimeVul-CoT originally outputs a binary label, so we use its explanation-generating variant so that it can be evaluated on the same basis as the other approaches.
Another common concern is that the dataset itself may have been part of the LLMs' training data. To mitigate data leakage, we explicitly collect only vulnerabilities disclosed after the backbone model's training cutoff, so that the model is unlikely to have seen them during training.
Finally, the evaluation itself relies on an LLM-as-a-Judge, whose reliability we address under construct validity.

\textbf{External validity}.
Our benchmark contains 50 vulnerable functions, a size that may limit how far our findings generalize. To reduce bias, we sample randomly from recently disclosed CVEs collected by a third-party tool~\cite{dataset:cvefixes} and retain only those whose function contains the CVE's root cause, rather than selecting cases favorable to our approach. Despite its size, the benchmark is diverse and its type distribution closely tracks that of the 622-CVE pool it was sampled from (Figure~\ref{fig:typerecall}). Its 50 vulnerable functions come from 45 distinct projects and span 10 programming languages and 10 vulnerability types, covering both memory-related and semantic categories. We therefore believe it is reasonably representative of real-world vulnerabilities.

%% file: sections/0x70Conclusion.tex
\section{Conclusion}

Many real-world vulnerabilities are semantic and domain-specific, and cannot be captured by the syntactic patterns that traditional static analysis relies on. 
We propose \beliefanalyzer, a belief-guided semantic static analysis implemented with large language models. Given a function, \beliefanalyzer identifies the actions implemented by the code, infers the beliefs behind these actions, and checks whether each belief is unreasonable or can be violated under specific conditions, and finally reports potential vulnerabilities with explanations of their root causes and exploitation contexts.

To address the context-dependency problem of binary metrics in function-level vulnerability detection, we propose two metrics, recall and overhead. Recall measures detection effectiveness as the fraction of known vulnerabilities for which an approach produces a matching explanation. Overhead measures human inspection effort as the number of findings that must be examined per identified vulnerability.

Our evaluation on 50 recent real-world vulnerable functions shows that \beliefanalyzer achieves the highest recall among all evaluated approaches and the best recall-overhead trade-off, and is effective across both memory-related and semantic vulnerabilities. 